\documentclass[aps,prl,twocolumn,superscriptaddress,showpacs,amsmath,amssymb]{revtex4}

\usepackage{graphicx}
\usepackage{bm}
\usepackage{amssymb}

\begin{document}

\title{Relativistic Electron Vortex Beams: Angular Momentum and Spin-Orbit Interaction}

\author{Konstantin Y. Bliokh}
\affiliation{Applied Optics Group, School of Physics, National University of Ireland, Galway, Galway, Ireland}
\affiliation{Advanced Science Institute, RIKEN, Wako-shi, Saitama 351-0198, Japan}

\author{Mark R. Dennis}
\affiliation{H.~H.~Wills Physics Laboratory, University of Bristol, Bristol BS8 1TL, United Kingdom}

\author{Franco Nori}
\affiliation{Advanced Science Institute, RIKEN, Wako-shi, Saitama 351-0198, Japan}
\affiliation{Physics Department, University of Michigan, Ann Arbor, Michigan 48109-1040, USA}

\begin{abstract}
Motivated by the recent discovery of electron vortex beams carrying orbital angular momentum (AM), we construct exact Bessel-beam solutions of the Dirac equation. 
They describe relativistic and nonparaxial corrections to the scalar electron beams. 
We describe the spin and orbital AM of the electron with Berry-phase corrections and predict the intrinsic spin-orbit coupling in free space. 
This can be observed as a spin-dependent probability distribution of the focused electron vortex beams. 
Moreover, the magnetic moment is calculated, which shows different $g$-factors for spin and orbital AM and also contains the Berry-phase correction. 
\end{abstract}

\pacs{41.85.-p, 03.65.Pm, 42.50.Tx, 03.65.Vf}

\maketitle

\textit{Introduction.---}Vortex beams carrying quantized orbital AM (OAM) along their axes of propagation are widely studied and exploited in modern optics [1]. 
Some of us previously predicted the existence of free-space vortex beams for non-relativistic scalar electrons [2]. 
More recently, electron vortex beams have been produced experimentally by several groups [3--5], using spiral phase plates [3] and nanofabricated diffraction holograms [4,5] which allow generation of beams with an OAM up to 100$\hbar$ [5]. 
This opens a promising avenue in electron microscopy [4], can be employed in quantum problems involving the AM interaction with external fields [2] and other particles [6], and deepens the analogy between light and matter waves.

The above studies dealt with the simplest approximation of a scalar electron and paraxial wave equation. 
However, a self-consistent description of the AM properties of the electron requires the exact Dirac equation, which describes the spin AM (SAM) intrinsically, and takes relativistic effects into account. 
Indeed, in optics, the full vector Maxwell equations are necessary to describe nonparaxial vortex beams and the spin-orbit interaction (SOI) phenomena arising upon the focusing or scattering of vortex beams [7--11]. 
Although zero-order Dirac wave packets are studied [12] and higher-order modes are used in confined guiding potentials [13], the OAM vortex solutions for free relativistic electrons have not been previously considered.

In this Letter we construct exact Bessel-beam solutions representing the AM eigenstates of a free Dirac electron. 
Using the recent unifying description of the AM and SOI of electromagnetic waves [11] and the semiclassical theory of Dirac wave packets [14--16], we give a self-consistent description of the OAM and SAM properties of the Dirac electron, and predict a number of SOI phenomena which can be verified experimentally.

\textit{Bessel beams from the Dirac equation.---}We start with the Dirac equation [17] in units with $c=1$: 
\begin{equation}\label{eqn:1}
i\hbar \partial _t \psi  = \left( {{\bm \alpha } \cdot {\bf p} + \beta \,m} \right)\psi~, 
\end{equation}
where $\psi \left( {{\bf r},t} \right)$ is the Dirac bispinor, $\bm{\alpha}$ and $\beta$ are the $4\times 4$ Dirac matrices, ${\bf p} =  - i\hbar \partial _{\bf r}$ is the momentum operator, and $m$ is the electron mass. 
The positive-energy momentum eigenstates (plane waves) of Eq.~(1) are [17]
\begin{equation}\label{eqn:2}
\psi _p \left( {{\bf r},t} \right) = W\left( {\bf p} \right)\exp \left[ {i\hbar ^{ - 1} \left( {{\bf p} \cdot {\bf r} - E\,t} \right)} \right]~,
\end{equation}
where $E = \sqrt {p^2  + m^2 }$, and 
%
\begin{equation}\label{eqn:3}
W = {1 \over {\sqrt {2} }} 
\begin{pmatrix}
   {\sqrt{ 1 + \dfrac{m}{E} }\,w}  \\ 
   {\sqrt {1 - \dfrac{m}{E}} \,{\bm \sigma } \cdot {\bm \kappa }\,w}  
\end{pmatrix}~.
\end{equation}
Here ${\bm{\sigma}}$ are the Pauli matrices, $\bm{\kappa}={\bf p}/p$ is the momentum-direction vector, and $w=(\alpha,\beta)^T$, $w^\dag w = 1$, is the 2-component spinor characterizing the electron polarization in the rest frame with $E=m$. 
The probability density and current are defined as
\vspace*{-0.1cm}
\begin{equation}\label{eqn:4}
\rho  = \psi ^\dag  \psi~,~~{\bf j} = \psi ^\dag  {\bm \alpha }\psi~,
\end{equation}
so that $\rho_p =1$ and ${\bf j}_p = {\bf p}/E$ for the plane wave (2) and (3). 
The basic states of the polarization, $w$, can be chosen as eigenstates of $\sigma_z$, i.e., the spin $z$-component in the rest frame [17]. 
Below we use such states $w^s$: $w^{1/2} =(1,0)^T$ and $w^{-1/2} =(0,1)^T$, with eigenvalues $s=\pm 1/2$.

Bessel beams [18] represent a superposition of monoenergetic ($p = \text{const}$) plane waves forming a fixed polar angle $\theta=\theta_0$ with the $z$-axis, Fig.~1a. 
They are uniformly distributed over the azimuthal angle $\phi\in (0,2\pi)$ with a vortex phase dependence $e^{i\ell\phi}$, $\ell =0,\pm 1,\pm 2,...$. 
Using cylindrical coordinates $(r,\varphi,z)$ in real space and $\left( {p_ \bot  ,\phi ,p_\parallel  } \right) = \left( {p\sin \theta ,\phi ,p\cos \theta } \right)$ in momentum space, we write the Fourier spectrum for the Bessel beam as:
\vspace*{-0.1cm}
\begin{equation}\label{eqn:5}
\tilde \psi _\ell  \left( {\bf p} \right) = {1 \over {i^\ell  p_{ \bot 0} }}W\!\left( {\bf p} \right)\delta\!\left( {p_ \bot   - p_{ \bot 0} } \right)e^{i\ell \phi }~,
\end{equation}
where $p_{ \bot 0}  = p\sin \theta _0$, $p_{\parallel 0}  = p\cos \theta _0$. 
The beam field is given by the Fourier integral of $\tilde \psi _\ell  \left( {\bf p} \right)$, which yields
\vspace*{-0.1cm}
\begin{equation}\label{eqn:6}
\psi _\ell  \left( {{\bf r},t} \right) = {{e^{i\Phi } } \over {2\pi i^\ell  }}\int\limits_0^{2\pi } {W\left( {\bf p} \right) e^{\left[ {i\xi \cos \left( {\phi  - \varphi } \right) + i\ell \phi } \right]}} \,d\phi~,
\end{equation}
where $\Phi \left( {z,t} \right) = \hbar ^{ - 1} \left( {p_{\parallel 0} z - E\,t} \right)$ and $\xi  =  k_{ \bot 0} r$, $k_{ \bot 0}=p_{ \bot 0}/\hbar$. 
Substituting here Eq.~(3) and assuming that the polarization amplitudes $w = \left( {\alpha ,\beta } \right)^T$ are the same for all the plane waves, we evaluate the integral (6) resulting in
\begin{widetext}
\vspace*{-0.3cm}
\begin{equation}
\label{eqn:7}
\psi _\ell   = {{e^{i\Phi } } \over {\sqrt {2} }}
\left[  
\begin{pmatrix}
   {\sqrt {1 + \dfrac{m}{E}} \,w}  \\ 
   {\sqrt {1 - \dfrac{m}{E}} \,\sigma _z \cos \theta _0 \,w} 
\end{pmatrix}
 e^{i\ell \varphi } J_\ell  \left( \xi  \right) + i
\begin{pmatrix}
   0  \\ 
   0  \\ 
   {-\beta \sqrt {\Delta } }  \\ 
   0   
\end{pmatrix}
 e^{i\left( {\ell  - 1} \right)\varphi } J_{\ell  - 1} \left( \xi  \right) + i
\begin{pmatrix}
   0  \\ 
   0  \\ 
   0  \\ 
   {\alpha \sqrt {\Delta } } 
\end{pmatrix}
e^{i\left( {\ell  + 1} \right)\varphi } J_{\ell  + 1} \left( \xi  \right) 
\right]~,
\end{equation}
\end{widetext}
where $\Delta  = \left( {1 - \dfrac{m}{E}} \right)\sin ^2 \theta _0$.

\begin{figure}[t]
\includegraphics[width=8.4cm, keepaspectratio]{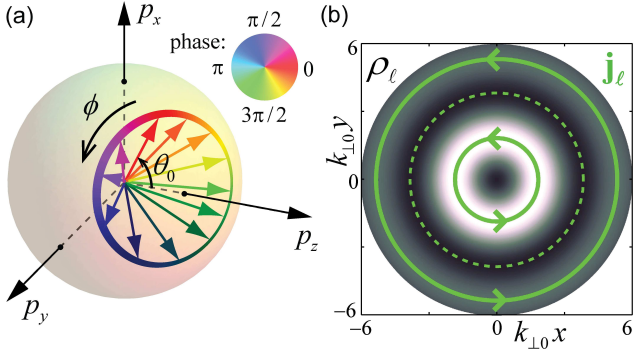}
\caption{(Color online) (a) The Bessel-beam spectrum (5) forms a cone of plane waves with a fixed polar angle $\theta=\theta_0$ and color-coded azimuthal phase $e^{i\ell\phi}$ on the monoenergetic sphere $p=\text{const}$. 
(b) An example of the probability density and azimuthal current distributions for the scalar Bessel beam with $\ell=1$ 
(the dashed line shows the zero of the current).} \label{fig1}
\end{figure}

Equation (7) describes \textit{spinor Bessel beams for Dirac electrons}, i.e., the electron counterpart of optical vector Bessel beams [11,19]. 
Hereafter, we use modes $\psi _\ell ^s$ with polarizations $w=w^s$, $s=\pm 1/2$. 
The first term in the square brackets represents a scalar-like Bessel beam of the order of $\ell$: $\psi _\ell   \propto J_\ell ^{} \left( \xi  \right)e^{i\left( {\ell \varphi + \Phi} \right)}$, Fig.~1b, whereas the terms proportional to $\sqrt{\Delta}$ describe the polarization-dependent coupling to other modes of the order of $(\ell + 2s)$. 
Thus, the latter terms describe the \textit{intrinsic SOI}, determined by the strength $\Delta$, which vanishes both in the paraxial ($\theta _0  \to 0$) and non-relativistic ($p \to 0$) limits as $\Delta  \approx \theta _0^2 p/m$. Using the canonical OAM and SAM operators,
%
\begin{equation}\label{eqn:8}
{\bf L} = {\bf r} \times {\bf p}~,~~\bm{\Sigma}=\frac{\hbar}{2}\text{diag}(\bm{\sigma},\bm{\sigma})~,
\end{equation}
one can readily see that in the limit $\Delta \to 0$, the solutions (7) are eigenmodes of both $L_z  =  - i\hbar \partial _\varphi$ and $\Sigma_z$, with eigenvalues $\ell$ and $s$, respectively. 
However, in the general nonparaxial relativistic case $\Delta \neq 0$, the Bessel beams $\psi_{\ell}^{s}$ are eigenmodes of the total AM $J_z =L_z +\Sigma_z$: $J_z \psi _\ell ^ s   = \left( {\ell  + s} \right)\psi _\ell ^s$, but \textit{not} of the OAM and SAM separately, cf. [11,19]. 
This also demonstrates the SOI.

The intrinsic SOI of a free Dirac electron manifests itself in the \textit{spin-dependent probability density} and current distributions, cf.~[11]. 
Substituting the spinorial Bessel beam $\psi_{\ell}^{s}$, Eq.~(7), into Eq.~(4), we obtain
%
\begin{eqnarray}\label{eqn:9}
\rho _\ell ^ s  \left( \xi \right) &=& \left( {1 - \frac{\Delta}{2}} \right)J_\ell ^2 \left( \xi  \right) + \frac{\Delta}{2} \,J_{\ell  + 2s}^2 \left( \xi  \right)~,\\
{\bf j}_\ell ^ s \left( \xi \right) &=& {{p_{\parallel 0}\over E} J_\ell ^2 \left( \xi  \right)} \, {\bf \hat{z}} + 
{{p_{ \bot 0}\over E} J_\ell  \left( \xi  \right)J_{\ell  + 2s} \left( \xi  \right)} \,{\bm{\hat{\varphi}}}~,
\end{eqnarray}
with ${\bm{\hat{\varphi}}}$ and ${\bf \hat{z}}$ being the corresponding unit vectors. 
The distributions $\rho _\ell ^ s$ and $|{\bf j}_\ell ^ s|$ are invariant w.r.t.~the transformation $(\ell,s)\to (-\ell,-s)$, but neither w.r.t.~$(\ell,s)\to (\ell,-s)$ nor $(\ell,s)\to (-\ell,s)$, which reflects the $\ell\cdot s$ symmetry typical of SOI. 
Figure 2 shows the fine \textit{spin-dependent splitting} of the density profiles $\rho _\ell ^ s  \left( \xi  \right)$ originated from the second SOI term in Eq.~(9). 
In the important particular case $|\ell |= 1$, the SOI term has A radically different behaviour at $\xi =0$ for the states $2s=\ell$ and $2s=-\ell$, and the effect is \textit{drastically enhanced}. 
Then the SOI term is proportional to $J_0 ^2 \left( \xi  \right)$ and $J_2 ^2 \left( \xi  \right)$ for the two states, resulting in, respectively, a \textit{finite} (despite the vortex!) and a zero intensity in the beam center, Fig.~2b. 
This effect can be observed for reasonable experimental parameters, akin to the tight focusing of circularly polarized optical beams with $|\ell |= 1$ [8,9,20]. 
The azimuthal current $({\bm j}_{\ell}^{s})_{\varphi}$, Eq.~(10), exhibits similar behaviour depending on the $(\ell,s)$ quantum numbers, Fig.~2.

\begin{figure}[t]
\includegraphics[width=8.8cm, keepaspectratio]{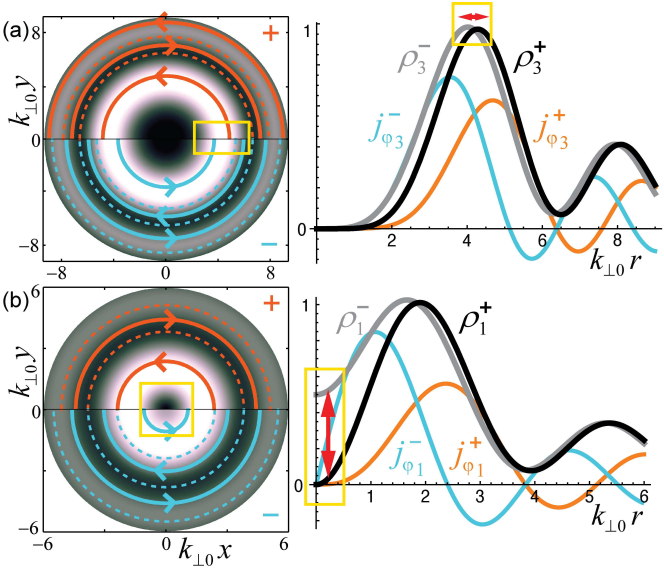}
\caption{(Color online) Spin-dependent distributions of the probability density (9) and azimuthal current (10) for the nonparaxial Bessel beams with $\ell=3$ (a) and $\ell=1$ (b). 
The $s=\pm 1/2$ spin states are indicated by ``+'' and ``--'' signs, and the SOI parameter is $\Delta=0.3$, which corresponds to $p\simeq 2.4 m$ and $\theta_0 = \pi/4$. 
The drastic difference between $\rho_1^+ (0)$ and $\rho_1^- (0)$ should be experimentally observable, cf. [9,20].} \label{fig2}
\end{figure}

\textit{Semiclassical theory of the electron AM.---}The expectation values of the OAM and SAM for the Bessel beams (7) can be determined using canonical operators (8). 
However, to reveal general features of the AM of the Dirac electrons, it is useful to consider the Foldy-Wouthuysen (FW) momentum representation [17,14,15], separating the positive- and negative-energy components in the Dirac equation. 
The FW transformation is a unitary analogue of the Lorentz boost to the rest frame: $\psi  = U_{FW}\left( {\bf p} \right) \psi '$,
%
\begin{equation}\label{eqn:11}
U_{FW}  = {1 \over {\sqrt {2} }}\left( {\sqrt {1 + \frac{m}{E}}  - \beta\, {\bm \alpha } \cdot {\bm \kappa }\sqrt {1 - \frac{m}{E}} } \right)~.
\end{equation}
It diagonalizes the Dirac Hamiltonian (1): $U_{FW}^\dag  \left( {{\bm \alpha } \cdot {\bf p} + \beta \,m} \right)U_{FW}  = \beta E$, and yields $W' = U_{FW}^\dag  W = \left(w,0 \right)^T$ for the plane waves (3). 
Owing to the momentum dependence, calculations of the coordinate-dependent quantities in the FW representation brings about additional gauge fields induced by the transformation (11) [14,15]. 
Calculating the operators of the observables ${\bf O} = \left\{ {{\bf r},{\bf L},{\bf S}} \right\}$ in the FW representation and using projection onto the positive-energy subspace $\wp ^ +$ ($2\times 2$ upper left sector of the matrix operators), we determine the effective operators for electron states, which exclude the influence of the negative-energy levels: $\bm{\mathcal O} = \wp ^ +  \left( {U_{FW}^\dag  {\bf O}\,U_{FW} } \right)$ [14,15,21].

In this manner, the operator for the electron position reads [14--16]:
%
\begin{equation}\label{eqn:12}
\bm{\mathcal{R}} = {\bf r} + \hbar \bm{\mathcal A}~,~~\bm{\mathcal A} = {{{\bf p} \times {\bm{\sigma} }} \over {2p^2 }}\left( {1 - {m \over E}} \right)~.
\end{equation}
Here $\bm{\mathcal A}=i \wp^+ (U_{FW}^\dag \partial_{\bf p} U_{FW})$ is the non-Abelian \textit{Berry connection} (gauge-field), which brings about the effective space noncommutativity, $\left[ {\mathcal R}^i ,{\mathcal R}^j \right] = i\hbar ^2 \varepsilon _{ijk} {\mathcal F}^k$, with the corresponding Berry curvature [14--16]
%
\begin{equation}\label{eqn:13}
\bm{\mathcal F} =  - {1 \over {2E^2 }}\left[ {{m \over E}{\bm \sigma } + \left( {1 - {m \over E}} \right){\bm \kappa }\left( {{\bm \kappa } \cdot {\bm \sigma }} \right)} \right]~.
\end{equation}
In a similar way, we derive the OAM and SAM operators which may be written as (cf.~[11]):
\begin{eqnarray}\label{eqn:14}
\bm{\mathcal L} &=& {\bf L} + \bm{\Delta} = \bm{\mathcal R} \times {\bf p}~,\\
\bm{\mathcal S} &=& {\bf S} - \bm{\Delta} = {m \over E}{\bf S} + \left( {1 - {m \over E}} \right){\bm \kappa }\left( {{\bm \kappa } \cdot {\bf S}} \right)~.
\end{eqnarray}
Here ${\bf S} = \wp^{+}({\bm \Sigma}) = \hbar {\bm \sigma }/2$ is the non-relativistic electron spin operator and 
\begin{equation}\label{eqn:16}
{\bm \Delta } = \hbar \bm{\mathcal A} \times {\bf p} =  - \left( {1 - \frac{m}{E}} \right){\bm \kappa } \times \left( {{\bm \kappa } \times {\bf S}} \right)
\end{equation}
is the \textit{SOI operator} describing the AM part effectively transferred from the SAM to the OAM: $\bm{\mathcal L} + \bm{\mathcal S} = {\bf L} + {\bf S}$.

The operators $\bm{\mathcal R}$, $\bm{\mathcal L}$, and $\bm{\mathcal S}$ describe the corresponding observables for the positive-energy electron states. 
Similar operators were discussed by Pryce in [22]. 
In the relativistic limit $m/E \to 0$, Eqs.~(12)--(16) are quite similar to the results obtained in [11] for photons, whereas the expression (15) is similar to the expectation value of the spin of an electron wave packet derived in [16]. 
Note that the operator $\bm{\mathcal S}$ yields the helicity operator ${\bm \kappa }\left( {{\bm \kappa } \cdot {\bf S}} \right)$ in the relativistic limit and the spin operator ${\bf S}$ in the nonrelativistic limit $(1-m/E) \to 0$.

Using the above operators, we calculate the normalized expectation values of observables for the Bessel beams: $\left\langle {\bf O} \right\rangle  = \left\langle \tilde{\psi}_\ell ^s  \right|\bm{\mathcal O}\left| \tilde{\psi}_\ell ^s  \right\rangle /\left\langle \tilde{\psi}_\ell ^s  \right|\left. \tilde{\psi}_\ell ^s  \right\rangle$. 
Here the spectrum (5) in the FW momentum representation reads $\left| \tilde{\psi}_\ell ^s \right\rangle \propto w^s \delta \left( {p_ \bot   - p_{ \bot 0} } \right)e^{i\ell \phi }$ and the convolution means 2D integration in momentum space [11]. As a result we arrive at [23] $\left\langle {\bf R} \right\rangle  = z\,{\bf \hat z}$, $\left\langle {\bf P} \right\rangle  = p_{\parallel 0} {\bf \hat z}$,
\begin{equation}\label{eqn:17}
\left\langle {\bf L} \right\rangle  = \hbar \left( {\ell  + \Delta \,s} \right){\bf \hat z}~,~~
\left\langle {\bf S} \right\rangle  = \hbar \left( {s - \Delta \,s} \right){\bf \hat z}~.
\end{equation}
Thus, the expectation values (17) are obtained from the operators (14)--(16) via the simple mapping: ${\bf L} \to \hbar \ell \,{\bf \hat z}$, ${\bf S} \to \hbar s \,{\bf \hat z}$, and ${\bm \Delta} \to \hbar s\, \Delta\, {\bf \hat z}$.

Equations (17) evidence a \textit{spin-to-orbit AM conversion} for nonparaxial relativistic electron beams (cf.~the AM conversion in optics [7--11]). 
This is a manifestation of the SOI intimately related to the spin-dependent probability density (9) (Fig.~2). 
Indeed, using ${\bf S} \to \hbar s \,{\bf \hat z}$ in the Berry connection (12) and integrating it along the contour of the Bessel-beam spectrum in momentum space (Fig.~1a), we obtain the \textit{Berry phase} $\Phi_B=-\oint{\bm{\mathcal A}}\cdot d{\bf p}=2\pi \Delta \, s$. 
The $(\ell,s)$-dependent radius of the beam is effectively described by the \textit{quantization condition} for the cylindrical caustic: $k_{\bot 0} R_{\ell}^{s}= \ell + \Delta\, s=\langle L_z \rangle / \hbar$, which arises from the total phase $2\pi\ell + \Phi_B$ around the beam [11]. 
Thus, the spin-dependent parts of the OAM (17) and probability density (9) both originate from the Berry phase gained after traversing the circular beam spectrum.

\textit{Magnetic moment.---}
The interaction of localized electron states with an external magnetic field is characterized by the magnetic moment originating from the AM. The one-electron magnetic moment can be calculated as
\begin{equation}\label{eqn:18}
\langle {\bf M} \rangle = {e \over 2}{{\int {\left({\bf r} \times {\bf j}\right)} \,dV} \mathord{\left/
 {\vphantom {{\int {{\bf r} \times {\bf j}} \,dV} {\int \rho  \,dV}}} \right.
 \kern-\nulldelimiterspace} {\int \rho  \,dV}}~,
\end{equation}
where $e<0$ is the electron charge. Using the definition of the probability current (4), Eq.~(18) can be represented as the expectation value $\langle {\bf M} \rangle = \left\langle \psi  \right|{\bf M}\left| \psi  \right\rangle /\left\langle \psi  \right|\left. \psi  \right\rangle$ of the operator ${\bf M} = {e \over 2}{\bf r} \times {\bm \alpha }$ in the coordinate representation. 
(This operator resembles the OAM (8) but with the \textit{velocity} operator ${\bm \alpha }$ instead of the momentum ${\bf p}$.) 
Using this analogy, one can calculate the effective magnetic-moment operator in the FW momentum representation for positive-energy electron: $\bm{\mathcal M} = \wp ^ +  \left( {U_{FW}^\dag  {\bf M}\,U_{FW} } \right)$. 
Cumbersome but straightforward calculations yield
\begin{equation}\label{eqn:19}
\bm{\mathcal M} = {e \over 2E} \left(\bm{\mathcal L}+2\bm{\mathcal S} \right)~.
\end{equation}
Hence, using Eqs.~(17), the magnetic moment for the Bessel-beam state is
\begin{equation}\label{eqn:20}
\langle {\bf M} \rangle = {e\hbar \over 2E} \left( \ell + 2s - \Delta\, s \right) {\bf \hat z}~.
\end{equation}
Thus, in complete agreement with our suggestion in [2], the magnetic moment is characterized by a $g=1$ factor for the OAM and a $g=2$ factor for the SAM of electron vortex states. 
At the same time, the SOI Berry-phase correction in Eq.~(20) disagrees with the previous wave-packet calculations [15,16] for the case $\ell =0$. 
This is explained by the fact that previous approaches neglected $\left\langle {\bf L} \right\rangle$, which does not vanish even at $\ell =0$ due to nonparaxial corrections. The magnetic moment (20) is an important observable that manifests itself in the Zeeman interaction with an external magnetic field.

\textit{Discussion.---}The above operator calculations of the expectation values imply a one-electron state, i.e., in fact, a wavepacket state of finite extent in the propagation direction [2]. 
For $z$-independent infinite beams, the number of electrons diverges, and one can use the \textit{linear densities} of the characteristics per unit $z$-length [1,7]. 
We calculated the linear densities of the OAM, SAM, and magnetic moment per one electron per unit $z$-length: $\langle \bar{L}_z \rangle$, $\langle \bar{S}_z \rangle$, and $\langle \bar{M}_z \rangle$. 
To regularize the corresponding integrals over the Bessel-beam cross-section, we modulate the transverse electron amplitude by a Gaussian of width $a$, and then consider the limit $a \to \infty$. Analytic calculations for such electron Bessel--Gauss beams [24] give:
\begin{equation}\label{eqn:21}
\langle \bar{L}_z \rangle = \hbar \left( {\ell  + \Delta \,s} \right)~,~~
\langle \bar{S}_z \rangle = \hbar s~,~~\langle \bar{M}_z \rangle = \ell + 2s~.
\end{equation}
This coincides with Eqs.~(17) and (20) in the paraxial limit but differs by terms of order $\Delta$. Comparative analysis of the one-particle expectation values and linear densities in the nonparaxial beams deserves further consideration and will be published elsewhere.

To conclude, we have constructed the exact cylindrical solutions of the Dirac equation, which exhibit nontrivial spin- and vortex-dependent properties. The intrinsic SOI phenomena originate from the Berry-phase terms and can be observed experimentally. 
We have calculated the observable OAM, SAM, and magnetic moment for exact relativistic electron vortex states. 

We acknowledge support from the European Commission (Marie Curie Action), Science Foundation Ireland (Grant No. 07/IN.1/I906), the Royal Society, LPS, NSA, ARO, DARPA, AFOSR, NSF grant No. 0726909, JSPS-RFBR contract No.~09-02-92114, Grant-in-Aid for Scientific Research (S), MEXT Kakenhi on Quantum Cybernetics, and the JSPS through its FIRST program.



\begin{thebibliography}{99}
\bibitem{Normalization}Despite $\left\langle \tilde{\psi}_\ell ^s \right|\left. \tilde{\psi}_\ell ^s \right\rangle  = \infty$ for Bessel beams, the normalized expectation values are finite. To show this, one has to use the rigorous definition of a Dirac delta distribution as a limit of a sequence of regular (e.g., Gaussian) functions.
\end{thebibliography}
\end{document}